\documentclass[preprint,showpacs,nofootinbib,prc,aps,english,unsortedaddress]{revtex4}
\usepackage[T1]{fontenc}
\usepackage{graphicx}
\usepackage{epsfig}
\usepackage{amsmath}
\usepackage{amssymb}
\usepackage{latexsym}

\usepackage[usenames]{color}

\begin{document}

\title{Kaon and pion femtoscopy at the highest energies available at the BNL Relativistic Heavy Ion Collider (RHIC) in a hydrokinetic model}

\author{Iu.A. Karpenko, Yu.M. Sinyukov}
\affiliation {Bogolyubov Institute for Theoretical Physics,
Metrolohichna str. 14b, 03680 Kiev,   Ukraine}

\begin{abstract}
The hydrokinetic approach, that incorporates hydrodynamic expansion
of the systems formed in A+A collisions and their dynamical
decoupling, is applied to restore the initial conditions and
space-time picture of the matter evolution in central Au+Au
collisions at the top RHIC energy. The analysis is based on the
detailed reproduction of the pion and kaon momentum spectra and
femtoscopic data in whole interval of the transverse momenta studied
by both STAR and PHENIX collaborations. The fitting procedure
utilizes the two parameters: the maximal energy density at supposed
thermalization time 1 fm/c and the strength of the pre-thermal flows
developed to this time. The quark-gluon plasma and hadronic gas is
supposed to be in complete local equilibrium above the chemical
freeze-out temperature $T_{ch}$ = 165 MeV with the equation of
states (EoS) at high temperatures as in the lattice QCD. Below
$T_{ch}$ the EoS in the expanding and gradually decoupling fluid
depends on the composition of the hadron-resonance gas at each
space-time point and accounts for decays of resonances into the
non-equilibrated medium. A good description of the pion and kaon
transverse momentum spectra and interferometry radii is reached at
both used initial energy density profiles motivated by the Glauber
and Color Glass Condensate (CGC) models, however, at different
initial energy densities. The discussion as for the approximate pion
and kaon $m_T$-scaling for the interferometry radii is based on a
comparison of the emission functions for these particles.
\end{abstract}

\pacs{25.75.-q, 25.75.Cj, 25.75.Ld }

\maketitle

\section{Introduction}
At present the majority of the dynamical models of A+A collisions,
which describe the soft physics phenomena are based on the Landau's
idea \cite{Landau} of space-time evolution of the thermal matter
formed in the collisions. This approach implies  at once the
specific space-time scales in the problem of nuclear scattering such
as a time of expansion, a volume occupied by the fireball,
hydrodynamic lengths, etc. The only direct tool to measure these
femtoscopic scales is the intensity interferometry method, often
called now as the femtoscopy. The measured scales - the
interferometry, or HBT radii - are associated just with the
homogeneity lengths in the rapidly expanding system created in heavy
ion collisions \cite{Sin}. So, the comparison with experimental data
of the space-time scales characterizing such and such dynamical
model of the system evolution and particle production should be one
of the first task of a justification of the model and discrimination
between different approaches. Nevertheless, such a comparison was
often ignored since almost all dynamic models, which pretend to be
complete and therefore describe the evolution of the matter as well
as its gradual decay, e.g., the hybrid (hydrodynamic plus UrQMD)
models \cite{hybrid} failed to reproduce pion out-, side-, long-
interferometry radii simultaneously with the hadronic spectra at
RHIC. Until now it was possible to reach only when some artificial
parametrization of freeze-out processes, e.g. a sudden freeze-out at
a fairly large temperature close to the hadronization one
\cite{Broniowski}, is utilized.

   In Refs. \cite{PRL,PRC, YadFiz} the HydroKinetic Model (HKM)
for  A+A collisions has been developed. It combines the advantages
of the hydrodynamic approximation, where possible phase transitions
are encoded in the corresponding equation of state (EoS), and
microscopic approach, accounting for a non-equilibrated process of
the spectra formation due to gradual particle liberation. The
dynamical decoupling is described by the particle escape
probabilities in inhomogeneous hydrodynamically expanding systems in
the way consistent with the kinetic equations in the relaxation time
approximation for emission function \cite{PRL}. The method can be
applied to match correctly hydrodynamics and UrQMD using as the
input the locally non-equilibrated distribution functions from the
HKM. Then one can match these models at space-like hypersurfaces
related to the late stage of the evolution, escaping thus the
inconsistencies connected with an inapplicability of hadron cascade
models at very high densities and with the causality \cite{Bugaev1}.

  The HKM method also allows one to take
into account a back reaction of particle emission on the
hydrodynamic evolution that corresponds an account of the viscous
effects at the hadronic stage of the evolution \cite{PRC}. It is
worth noting that found in the HKM ratio of the shear viscosity to
the total entropy is less than 1/2 in the space-time region of
maximal hadronic emission \cite{Noneq}. An analysis of the QGP
evolution within viscous hydrodynamics is also a topical problem
since the shear viscosity brings an important effect, an increase of
transverse flows during the evolution \cite{Teaney}. However, until
the viscosity of the QGP as the function of the temperature becomes
clear, this effect is simpler to take into account in the
phenomenological way, as it is proposed in what follows.

In this article we apply the HydroKinetic Model (HKM) \cite{PRL,PRC}
to an analysis of the femtoscopic measurements at RHIC for central
Au+Au collisions at the top energy $\sqrt{s}=200$ AGeV. Namely, we
analyze pion and kaon transverse momentum spectra and the $m_T$-
behavior of the pion and kaon interferometry radii to clarify, in
particularly, how these observables depend on the initial
conditions: Glauber and CGC-like. The basic hydrokinetic code,
proposed in \cite{PRC}, is modified now to include decays of
resonances into the expanding hadronic chemically non-equilibrated
system and, based on the resulting composition of the
hadron-resonance gas at each space-time point, to calculate the
equation of state (EoS) in a vicinity of this point. The obtained
local EoS allows one to determine the further evolution of the
considered fluid elements. In the zone of chemical equilibrium,
above the chemical freeze-out temperature, the EoS is taken in
accordance with the lattice QCD results.

The paper is organized as the following. Section II is devoted to
the initial conditions (IC) for thermal evolution of the matter in
Au+Au collisions at RHIC. In Section III we discuss the EoS  of the
matter in equilibrated and chemically non-equilibrated zones. The
kinetics of the system in the non-equilibrium zone related to
system's evolution and decoupling is described in Section IV. The
underlying hydrodynamic model for both chemically equilibrated and
non-equilibrated domains is presented in Section V. Section VI is
devoted to the results obtained and discussions. The conclusions and
outlook are done in Section VII.

\section{Initial conditions for hydro-evolution of thermal matter}
 Our results are all related to the central rapidity slice where we
 use the boost-invariant Bjorken-like initial condition in longitudinal direction.
 We consider the proper time of thermalization of quark-gluon matter as the
 minimal one discussed in the literature, $\tau_0=1$ fm/c
 \cite{Zxu-1}.

 \subsection {Pre-thermal flows}
 If one starts the
 hydrodynamic evolution at the "conventional time"  $\tau_i=$1 fm/c  {\it without}
 transverse flow -
since no pressure is established before thermalization - the
resulting radial flow will not be developed enough to describe
simultaneously the absolute values of pion, kaon and proton spectra,
as well as the anisotropy of elliptic flow in non-cental collisions.
To describe the observables one needs to start the hydro-evolution
at very small initial time, $\tau \sim 0.5$ fm/c \cite{Heinz}, where
it is difficult to expect the thermalization. This controversial
situation is overcome due to the results of Ref. \cite {sin1} where
is shown that the initial transverse flows in thermal matter as well
as their anisotropy, leading to asymmetry of the transverse momentum
spectra in non-central collisions, could be developed at the
pre-thermal, either classical field (Glasma) \cite{Glasma}, string
\cite{string} or partonic stages, with even more efficiency than in
the case of very early hydrodynamics. So, the hypotheses of early
thermalization at times less than 1 fm/c is not necessary: the
radial and elliptic flows develop no matter whether a pressure
already established. The general reason for them is an essential
finiteness of the system in transverse direction. Then the flows of
particle number or energy directed outward the system cannot be
compensated by the inward directed (from periphery to the centre)
flows. This difference means the non-zero net flows no matter how
the collective velocity is defined: according to Ekkart or to
Landau-Lifshitz. The further development and exploitation of these
results were done in Refs. \cite{flow, APP, Sin-2}.

 The
initial transverse rapidity profile is supposed to be linear in
radius $r_T$:
\begin{equation}
y_T=\alpha\frac{r_T}{R_T},\quad \text{where}\quad R_T=\sqrt{<r_T^2>}
 \label{yT},
\end{equation}
here $\alpha$ is the second fitting parameter. Note that the fitting
parameter $\alpha$ should include also a positive correction for
underestimated resulting transverse flow since in this work we did
not account in direct way for the viscosity effects \cite{Teaney}
neither at QGP stage nor at hadronic one. In formalism of HKM
\cite{PRC} the viscosity effects at hadronic stage are incorporated
in the mechanisms of the back reaction of particle emission on
hydrodynamic evolution which we ignore in current calculations.
Since the corrections to transverse flows which depend on unknown
viscosity coefficients are unknown, we use fitting parameter
$\alpha$ to describe the "additional unknown portions" of flows,
caused both factors: by a developing of the pre-thermal flows and
the viscosity effects in quark-gluon plasma.

\subsection{Glauber-like initial transverse profile} A simple Glauber model
initialization assumes that the initial energy density in the
transverse plane is proportional to the participant nucleon density
\cite{Kolb},
\begin{equation}
\epsilon({\bf b},{\bf x}_T)=\epsilon_0 \frac{\rho({\bf b},{\bf
x}_T)}{\rho_0}
\label{GIC}
\end{equation}
with $\rho_0\equiv\rho(0,0)$ and
\begin{eqnarray}
\rho({\bf b},{\bf x}_T)=(T({\bf x}_T+{\bf b}/2) S({\bf x}_T-{\bf
b}/2)+T({\bf x}_T-{\bf b}/2)S({\bf x}_T+{\bf b}/2)),
\nonumber\\
S({\bf x}_T)=\left[1-\left(1-\sigma_{NN} \frac{T({\bf
x}_T)}{A}\right)^A\right], \label{rho}
\end{eqnarray}
where $A$ is atomic number, equal to 197 for Au+Au collision, and
$\sigma_{NN}=51$~mb(=5.1~fm${}^2$) is the nucleon-nucleon
cross-section at $\sqrt{s_{NN}}=200$~AGeV. The impact parameter
${\bf b}=(b,0)$ is equal to zero, $b$=0, in  the considered case of
central collision. The parameter $\epsilon_0\equiv \epsilon(b=0,{\bf
x}_T=0)$ is the maximal energy density at the initial moment of
thermalization. The thickness $T({\bf x}_T)$ is expressed through
the Woods-Saxon distribution profile:
\begin{equation}
T({\bf x}_T)=\int\limits_{-\infty}^\infty F_{\rm WS}({\bf x})dx_L,
\end{equation}
where
\begin{equation}
F_{\rm WS}({\bf x})=
\frac{a}{\exp{\left[\left.\left(\sqrt{x^2_L+x^2_T}-R_A\right)\right/\delta\right]}+1}.
\end{equation}
Here we use that $R_A=1.12A^{1/3}-0.86A^{-1/3}\approx6.37$~fm,
$\delta=0.54$~fm. Constant $a$ is obtained from normalization
condition:
\begin{equation}
\int T({\bf x}_T)d^2x_T=A.
\end{equation}

One can think that transversal Glauber-like $\epsilon$-profile has
been formed to some initial time $\tau_0\approx 0.1 - 0.3$ fm/c (see
below) when the system is not thermal yet.  However, the form of the
profile is, practically, not modified to supposed thermalization
time $\tau_0 \sim 1$ fm/c because the transverse velocities reached
to this time are relatively small. At the same time, the absolute
values of energy density can change significantly because of the
strong longitudinal expansion. We use the maximal energy density
$\epsilon_0$  at time $\tau_i=1$ fm/c as the second fitting
parameter.

\subsection{Initial conditions motivated by Color Glass Condensate
model}

 Within CGC effective field
theory some important physical properties of the field are defined
by the parameter $\Lambda_s=g^2\mu$ where $g^2=4\pi\alpha_s$ and
$\mu^2$ is dimensionless parameter, which is the variance of the
Gaussian weight over the color charges $\rho$ of partons. The value
of $\Lambda_{s0}$ is approximately equal to the saturation scale
value, $Q_s$, and for the RHIC energies one can use
$\Lambda_{s0}\approx Q_s\approx2$~GeV${}^2$ \cite{Lappi}. According
to the results of Refs.~\cite{KV,KNV}, (proper) time $\tau_0 \approx
3/\Lambda_s$ is an appropriate scale controlling the formation of
gluons with a physically well-defined energy. At later times the
dynamics of the classical Yang–-Mills fields produced in
nucleus-nucleus collisions can be linearized and approximated by
that of a system of weakly coupled harmonic oscillators. Then one
can compute the field amplitudes squared in momentum space and find
corresponding distribution for the gluon number~\cite{KNV,K} for
cylindrically homogeneous transverse profile. It has the form at
$p_T<1.5\Lambda_s$ and
$\eta=\frac{1}{2}\ln{\frac{t+x_L}{t-x_L}}\simeq0$,
\begin{eqnarray}
&&\frac{dN}{d^2p_Td^2x_Td\eta}\equiv f(T_{\rm eff})
\nonumber\\
&&=\frac{a_1}{g^2}\left[\exp{\left(\left.\sqrt{p_T^2+m_{\rm
eff}^2}\right/T_{\rm eff}\right)}-1\right]^{-1}, \quad
\label{distrib1}
\end{eqnarray}
where $m_{\rm eff}=a_2\Lambda_{s0}$, $T_{\rm eff}=a_3\Lambda_{s}$;
$a_2=0.0358$, $a_3=0.465$. The constant $a_1/g^2$ will be absorbed
into factor $\epsilon_0$  which is our fitting parameter.

The dependence of the distribution (\ref{distrib1}) on transverse
coordinates ${\bf x}_T$ is constructed as follows~\cite{KNV}:
\begin{equation}
\Lambda^2_s({\bf x}_T)=\Lambda^2_{s0} \frac{\rho({\bf b},{\bf
x}_T)}{\rho_0}. \label{Lambda}
\end{equation}
where the participant density at a particular position in the
transverse plane is defined by (\ref{rho}).

To define the initial energy density profile  we need the partonic
phase-space distribution $f{_0}(x,p)=dN/d^3xd^3p$. Note, that it is
associated with the hypersurfaces $t={\rm const}$. To express the
phase-space density through the values
$\frac{dN}{d^2x_Td^2pd{\eta}}$ defined at $\sqrt{t^2-x_L^2}=\tau_0$,
one should take into account that the density  of  partons with
momentum ${\bf p}$ crossing element $d^3\sigma(x)$ of this
hypersurface is
\begin{eqnarray}
&&p^0\left.\frac{dN}{d^3p}\right|_{d\sigma(x)}=
p^{\mu}d\sigma_{\mu}(x)f{_0}(x,p)
\nonumber\\
&&=f_0(x,p)\tau_0p_T\cosh{\theta}d^2x_Td\eta, \label{CF}
\end{eqnarray}
where $\theta=y-\eta$, $y$ is rapidity of partons (in momentum
space). Therefore
\begin{equation}
f_0(x,p)=\frac{1}{\tau_0
m_T\cosh{\theta}}\frac{dN}{d^2x_Td^2p_Td\eta dy}.
\label{phase-space}
\end{equation}

One can formally get the $d^6N$ distribution from (\ref{distrib1})
by multiplying it by $\delta$--function:
\begin{equation}
\frac{dN}{d^2p_Td^2x_Td\eta dy}=f(T_{\rm eff})\delta(y-\eta).
\label{GCG_density}
\end{equation}
Such a phase-space distribution, corresponding the CGC asymptotic
results \cite{KLW}, is widely used for a description of the initial
state in A+A collisions \cite{Grainer}. However, a presence of the
delta-function in the phase-space density contradicts evidently to
the basic principle of the quantum mechanics. Indeed, the classical
phase-space density has to follow from the quantum mechanical one in
some limit. The Wigner function $f_{\text{W}}(x,p)$ \cite{Wigner},
that is the quantum mechanical analog of the classical phase-space
density $f(x,p)$, satisfies the restriction $ \int
f_{\text{W}}^{2}(x,p)d^{3}pd^{3}x \leq (2 \pi \hbar)^{-3}$ (see e.g.
\cite{QM}, note that  the equality takes place for a pure state
only), here the normalization condition $\int
f_{\text{W}}(x,p)d^{3}pd^{3}x =1 $ is supposed. It evidently
excludes  utilization of the delta-function as factor in the
structure of the Wigner function. Therefore, in order to escape
contradiction with quantum mechanics, an another prescription,
instead of utilization of  delta function, should be used for the
longitudinal part of distribution $f(x,p)$; it can be, for example,
the boost-invariant prescriptions used in Ref. \cite{Sin-2}.
Following to this recept the smearing of $\delta$-function at
hypersurface $\tau_0$ in (\ref{GCG_density})  as follows
\begin{equation}
\frac{dN}{d^2p_Td^2x_Td\eta dy}=f\left(\frac{T_{\rm
eff}}{\cosh{(\eta-y)}}\right). \label{smearing}
\end{equation}
In this way we fix the phase-space density (\ref{phase-space}). This
may correspond to quasi-thermal averaged partonic distribution which
can be reached at moment $\tau_0$ due to quantum effects
(uncertainly principle), different kind of turbulences and
Schwinger-like mechanism of pair production in the pulse of strong
color field. It does not mean that the true thermalization which
should be supported by a permanent mechanism of partonic
interactions is reached at $\tau_0 \approx 3/\Lambda_s\approx 3$
fm/c.

As a result we use the following  form of boost-invariant
phase-space distribution for gluons at the initial hypersurface
$\tau_0$:
\begin{equation}
f_0=g^{-2}\frac{a_1(\tau_0m_T\cosh\theta)^{-1}}
{\exp{\left(\left.\sqrt{m_{\rm eff}^2({\bf
x}_T)+p^2_T}\cosh{\theta}\right/T_{\rm eff}({\bf x}_T)\right)}-1},
\label{basic-dist}
\end{equation}
 here $\theta = \eta - y$, ${\bf x}_T=(X,Y)=(x_T\cos\varphi,x_T\sin\varphi)$ and we consider gluons as massless
 particles, $m_T=p_T$.  Such a
 distribution depends on the effective mass $m_{\rm eff}({\bf x}_T)=a_2\Lambda_s({\bf x}_T)$
 and the temperature $T_{\rm eff}({\bf x}_T)=a_3\Lambda_s({\bf x}_T)$ (numerical values for
 $a_2$ and $a_3$ are the same as in Eq.~(\ref{distrib1})),  which, in accordance with
 Ref.~\cite{KNV}, are determined by the local scale $\Lambda_s({\bf x}_T)$ (\ref{Lambda}).

The components of the energy-momentum tensor in the pseudo-Cartesian
coordinates reads
\begin{equation}\label{TD}
T^{\mu\nu}(x)=\int p^\mu p^\nu f(x,p) p_Tdp_Tdyd\phi,
\end{equation}
 where the Lorentz-invariant integration measure $d^3p/p_0$ in the Cartesian variables
 is already re-written in Bj\"orken variables as $p_Tdp_Tdyd\phi$.

 We numerically calculate the components of the energy-momentum tensor with
 the distribution function, following from Eq.~(\ref{basic-dist}), at $\eta=0$.

 Note that, at $\tau=\tau_0$, the energy-momentum tensor takes the
 form
\begin{equation}
T^{\mu\nu}_0({\bf x}_T, x_L=0)=\frac{a_1}{g^2\tau_0}\Lambda^3_s({\bf
x}_T)t^{\mu\nu},
\end{equation}
 where $t^{\mu\nu}$ are the constant coefficients fixed by the constants $a_2$ and $a_3$.
 Therefore, the energy profile in transverse plane at $\tau_0$ in central collisions can
 be presented in the form (see (\ref{Lambda}))
 \begin{equation}
 \epsilon(x_T)=\epsilon_0\frac{\rho^{3/2}(0,x_T)}{\rho^{3/2}_0},
 \end{equation}
where the number of participants is defined by (\ref{rho}). Under
the same reason as for the Glauber-like IC  we use the form of this
profile to build the IC for hydrokinetic evolution at the
thermalization time $\tau_i=1 $ fm/c. The maximal energy density
$\epsilon_0$ at (proper) time $\tau_i$ is the fitting parameter as
in the case of the Glauber IC.

\section{The thermal matter in A+A collision and equation of state}
Here we  describe the matter properties and its thermodynamic
characteristics, e.g. equation of state, that are necessary
components of the hydrokinetic model. We suppose that soon after
thermalization the matter created in A+A collision at RHIC energies
is in the quark gluon plasma (QGP) state.
 Also at time $\tau_i$, there is a peripheral
region with relatively small initial energy densities: $\epsilon(r)<
0.5$ GeV/fm$^3$. This part of the matter ("corona") does not
transform into QGP and have no chance to be involved in
thermalization process \cite{Werner}. By itself the corona gives no
essential contribution to the hadron spectra \cite{Werner}. One
should consider it separately from the thermal bulk of the matter
and should not include in hydrodynamic evolution. Therefore we cut
the initial Glauber or CGC-like profiles at $\epsilon(r)\leq 0.5$
GeV/fm$^3$ when consider IC for hydrodynamic evolution of the
system.

During the system evolution the QGP is cooling and finally
transforms into hadron phase, most probably, according to the
crossover scenario. Such a transformation may occur in the interval
of the temperatures 170-190 MeV. At the temperature $T=T_{ch}\approx
165$ MeV the chemical freeze-out happens, as demonstrates an
analysis of the particle number ratios \cite{Becattini, PBM}. The
conception of the chemical freeze-out means that at the temperatures
$T\geq T_{ch}$ the bulk of the expanding matter is in the local
thermal and chemical equilibrium while at $T < T_{ch}$ the chemical
composition becomes in some sense frozen: one can neglect the
majority of inelastic reactions except for decays of resonances and
recombination processes. The hadronic matter in the later
thermodynamic region is not in the chemical equilibrium, moreover,
the  hadronic medium gradually emits particles being in this zone
and, so, loose, in addition, also the local thermal equilibrium.
Therefore, one should consider in different ways the matter
evolution in the two 4D space-time zones separated by the 3D
hypersurface corresponding to the isotherm $T=T_{ch}\approx 165$
MeV. Let us describe the thermodynamic  properties of matter in both
these regions.

\subsection{The EoS in the equilibrated space-time domain.}   At high
temperatures corresponding to the QGP phase and crossover transition
to hadron phase we use a realistic EoS \cite{Laine} adjusted to the
lattice QCD results for zero barionic chemical potential so that it
is matched with an ideal chemically equilibrated multicomponent
hadron resonance gas at $T_c=175$ MeV.
 To take into account a conservation of the net baryon number, electric charge and
 strangeness in the QGP phase, one has first to make corrections to
 thermodynamic quantities for nonzero chemical potentials.
 As it is proposed in \cite{karsch_nonzero}, a modification of the EoS
 can be evaluated by using of the Taylor series expansion in terms of the light and
  strange quark chemical potentials, or analogously in baryon and strange hadronic chemical potentials:
\begin{equation}
\frac{p(T,\mu_B,\mu_S)}{T^4}=\frac{p(T,0,0)}{T^4}+\frac 1 2
\frac{\chi_B}{T^2}\left(\frac{\mu_B}{T}\right)^2+ \frac 1 2
\frac{\chi_S}{T^2}\left(\frac{\mu_S}{T}\right)^2+\frac{\chi_{BS}}{T^2}\frac{\mu_B}{T}\frac{\mu_S}{T}
\label{eos-mu}
\end{equation}
The expansion coefficients $\chi_B$ and $\chi_S$ are the baryon
number and strangeness susceptibilities which are related to thermal
fluctuations of baryon number and strangeness in a thermal medium at
zero chemical potentials.

To obtain the EoS in the equilibrium zone we use the numerical
results for $\chi_B$ and $\chi_S$ as a function of the temperature
given in \cite{karsch_nonzero}. The values for the ratios $\mu_q/T$
in (\ref{eos-mu}) during the system evolution can be determined
approximately. If at some hypersurface corresponding to an isotherm,
like as at the chemical freeze-out hypersurface, the chemical
potentials are uniform, then the following ratios remain constants
$$
\frac{\mu_q}{T}=const_q,\quad \text{where}\quad q= B,S,E
$$
during the chemically equilibrated isoentropic evolution of the
Boltzmann massless gas. In our approximation we use there constrains
and find the corresponding constants from the chemical potentials
obtained together with $T_{ch}$ from an analysis of the particle
number ratios. In concrete calculations we use the chemical
freeze-out temperature $T_{ch}=165$ MeV, corresponding chemical
potentials $\mu_B$ =29 MeV, $\mu_S$ =7 MeV, $\mu_E$ =-1 MeV and also
the strangeness suppression factor $\gamma_S=0.935$ which are
dictated by 200A GeV RHIC particle number ratios analysis done in
the statistical model \cite{Becattini, PBM}.

\subsection{The EoS in the chemically non-equilibrated
domain.}

At the chemical freeze-out temperature $T_{ch}$ the  "lattice"
EoS
taken from \cite{Laine} and corrected for non-zero chemical
potentials is matched with good accuracy with ideal Boltzmann
hadronic resonance gas which includes $N=359$ hadron states made of
u, d, s quarks with masses up to 2.6 GeV.
Essentially, we use the same particle set in the FASTMC event generator
\cite{Amelin}.
Technically, in the numerical code, we input the corresponding N
functions  - the densities $n_i$ of each hadron $i$ and the
equations for $n_i$ already at the very beginning of the system
evolution; however, these densities are meaningless in the QGP phase
and their evaluation does not influence on the system evolution in
the equilibrated zone. These functions are brought into play at
$T<T_{ch}$. If this thermodynamic region would correspond to the
complete conservation of the particle numbers then, in addition to
the energy-momentum conservation, one would account for the
conservation equations for particle number flows in the form:
\begin{equation}\label{cfo}
    \partial_\mu(n_i u^\mu)=0, \quad i=1\dots N
\end{equation}

In our problem, however, during the system evolution in the
non-equilibrated zone $T<T_{ch}$ the resonance decays have to be
taken into account. The decay law in a homogeneous medium with $T\ll
m_i$ ($m_i$ is the resonances mass) implies a summing up of a
decrease of unstable $i$th particle number due to decays and an
increase because of decays of heavier $j$th resonance into $i$th
particle:
\begin{equation}\label{decay-law}
    \frac{dN_i}{dt}=-\Gamma_i N_i + \sum\limits_j b_{ij}\Gamma_j N_j
\end{equation}
where $\Gamma_i$ is the total width of resonance $i$,
$b_{ij}=B_{ij}M_{ij}$ denote the average number of $i$th particles
coming from arbitrary decay of $j$th resonance,
$B_{ij}=\Gamma_{ij}/\Gamma_{j,tot}$ is branching ratio, $M_{ij}$ is
a number of $i$th particles produced in $j\rightarrow i$ decay
channel. The set on N equations (\ref{decay-law}), solved together,
takes into account all possible cascade decays $i\rightarrow
j\rightarrow k\rightarrow\dots$. This also conserves net charges,
e.g. baryon, electric charge and strangeness, since the charges are
conserved in resonance decay process. If one relates the Eq.
(\ref{decay-law}) to the fluid element of some volume $\Delta V$
moving with four-velocity $u^{\mu}$, then a covariant relativistic
extension of the decay law for a hydrodynamic medium leads to the
equation (\ref{cfo}):
\begin{equation}
    \partial_\mu(n_i(x) u^\mu(x))=-\Gamma_i n_i(x) + \sum\limits_j b_{ij}\Gamma_j
    n_j(x)\label{decay}
\end{equation}
when one neglects a thermal motion of the resonance $j$, that can be
justified because  post (chemical) freeze-out temperatures are much
less than the mass of the lightest known resonance. Also, Eq.
(\ref{decay}) for the hydrodynamic evolution is written under
supposition of an instant thermalization of the decay products, that
is consistent with the ideal fluid  approximation (mean free path is
zero). In the kinetic part of the HKM we consider the next
approximation when the non-equilibrium character of the distribution
functions and the kinetics of resonance decays are taken into
account. We also can approximately  account for a recombination in
the processes of resonance decays into expanding medium just by
utilizing the effective decay width $\Gamma_{i,eff}=\gamma\Gamma_i$
in Eq. (\ref{decay}).  We use $\gamma = 0.75$ \cite{KarSin} for the
resonances containing $u$ and $d$ quarks supposing thus that about
30\% of such resonances are recombining during the evolution.

The equations (\ref{decay}) together with the hydrodynamic equations
and the equation of state should give one the energy density and
composition of the gas in each space-time points. To find the EoS
$p=p(\epsilon,\{n_i\})$ for the mixture of hadron gases we start
with the expressions for energy density and particle density for
$i$th component of multicomponent Boltzmann gas :
\begin{align}
\epsilon_i&={g_i\over 2\pi^2}m_i^2T(3TK_2(m_i/T)+mK_1(m_i/T))\exp(\mu_i/T) \nonumber\\
n_i&={g_i\over 2\pi^2}m_i^2 TK_2(m_i/T)\exp(\mu_i/T).
\end{align}
Then, the equation for the temperature is:
\begin{equation}\label{Tboltz}
    \epsilon=3nT+\sum\limits_i n_i m_i
    \frac{K_1(m_i/T)}{K_2(m_i/T)},
\end{equation}
where $n=\sum_i n_i$. Having solved this equation numerically for
given $\epsilon$ and $\{n_i\}$, we get the temperature and then find
the pressure using simple relation for multicomponent Boltzmann gas:
\begin{equation}\label{pboltz}
    p=nT
\end{equation}
The equations (\ref{Tboltz}), (\ref{pboltz}) define
$p=p(\epsilon,\{n_i\})$.

Thus, we follow the evolution of all N densities of hadron species
in hydro calculation, and compute EoS dynamically for each chemical
composition of N sorts of hadrons in every hydrodynamic cell in the
system during the evolution. Using this method, we do not limit
ourselves in chemically frozen or equilibrated evolution, keeping
nevertheless thermodynamically consistent scheme.

As it was mentioned before, we use the Boltzmann approximation in
the EoS calculation to decrease computational time. However, for
emission function and spectra calculation we use quantum
Bose-Einstein/Fermi-Dirac distribution functions with chemical
potentials calculated to give the same particle densities as in the
Boltzmann case.  We checked that the measure of relative divergence
in the energy density if one uses the quantum distribution functions
instead of the Boltzmann one, is not bigger than 3\% in the
thermodynamic region which is actually contributed to formation of
hadronic spectra.


\section{Kinetics in the non-equilibrium hadronic zone}
To describe the non-equilibrium evolution and decay of hadronic
system we start from the Boltzmann equations for the mixture of
hadrons, most of which have finite lifetimes and decay widths
compatible with particle masses. The set of such equations for
i-components of the hadron resonance gas which account for the only
binary interactions (elastic scattering) and resonance decays are:
\begin{equation}
 \frac{p^{\mu}_i}{p_i^0}\frac{\partial f_i(x,p)}{\partial x^{\mu }}
 =G_i^{scatt}-L_i^{scatt}(x,p)+ G_i^{decay}(x,p)-L_i^{decay}(x,p)\equiv G_i(x,p)-L_i(x,p).
 \label{boltz-}
\end{equation}

Here we ignore the processes of resonance recombination which
simpler to account phenomenologically (see the previous Section).
The term gain ($G$) describes an income of the particles into
phase-space point $(x,p)$ due to scatters and resonance decays. The
term {\it loss} ($L$) is related to a decrease of particles in the
vicinity of the phase space point $(x,p)$ due to re-scattering and
decays of resonances. The {\it loss} term is proportional to the
particle number density in the point $x$ and so
$L_i^{scatt}(x,p)=f_iR_i, \ L_i^{decay}(x,p)= f_iD_i$ where $R$ is
scattering rate, and $D$ is decay rate. If one considers the
equations for stable or quasi-stable particles, then
$L_i^{decay}(x,p)=0$ ($D_i \equiv 0$).

The method allowing to find the emission function of the hadrons
based on the Boltzmann equations in the (generalized) relaxation
time approximation was proposed in Refs. \cite{PRL,PRC}. Following
to this method we put:  $J_i(x,p)\approx R_{i,l.eq.}(x,p)$, $G_i
\approx R_{i,l.eq.}(x,p)f_{i,l.eq.}(x,p)+ G_i^{decay}(x,p)$. The
quantity  $R(x,p)=\tau^{-1}_{rel}(x,p)$ is the inverse relaxation
time, or collision rates in global reference frame. Then,
\begin{equation}
\frac{p^\mu}{p^0}\partial_\mu f_i(x,p) =
(f_{i}^{l.eq.}(x,p)-f_i(x,p))R_i(x,p)+G_i^{decay}(x,p)-L_i^{decay}(x,p)
\label{rel}
\end{equation}
The explicit form of $G_i^{decay}(x,p)$ term will be derived later.
In the first approximation to hydro-kinetic evolution the parameters
of the local equilibrium distribution function $f_{i,l.eq.}(x,p)$,
e.g. the temperature $T(x)$, chemical potentials $\mu_i(x)$ are
determined by the  hydrodynamic evolution. The details of
hydrodynamic approach used in the model are described in the next
section.

\subsection{Emission functions in hyperbolic coordinates and spectra formation} All our
results are related to the very central rapidity interval, $y
\approx 0$, and we will use the boost-invariant approach to describe
strong longitudinal matter expansion observed at RHIC. For such an
approach the hyperbolic coordinates in $(t,x_L)$ directions are more
suitable than the Cartesian ones. Then the kinetic equations take a
form
\begin{eqnarray}
\frac{1}{m_T\cosh
y}\left(m_T\cosh\theta\frac{\partial}{\partial\tau}-\frac{m_T\sinh\theta}{\tau}\frac{\partial
}{\partial\eta} +\vec p_T\frac{\partial }{\partial \vec
r_T}\right)f_i(\tau,\theta,{\bf r}_T,{\bf p}_T)=\nonumber \\
\left[f_{i}^{l.eq.}(\tau,\theta,{\bf r}_T,{\bf
p}_T)-f_i(\tau,\theta,{\bf r}_T,{\bf
p}_T)\right]R_i(\tau,\theta,{\bf r}_T,{\bf
p}_T)+G_i^{decay}(\tau,\theta,{\bf r}_T,{\bf p}_T) \label{BEcurved}
\end{eqnarray}
where $\tau=\sqrt{t^2-x_L^2}$ is a proper time,
$m_T=\sqrt{m^2+p_T^2}$ is a transverse mass, $\theta=\eta-y$, $\eta$
is a space-time rapidity, defined above Eq. (\ref{distrib1}), and
$y$ is a particle rapidity.

The formal solutions of (\ref{BEcurved}) correspond to the
non-equilibrium distribution functions in expanding and decaying
multi-hadronic system:
\begin{align}\label{f}
    &f_i(\tau,\theta,{\bf r}_T, {\bf p}_T)=f_i^{l.eq.}(\tau_0,\theta^{(\tau_0)}(\tau),{\bf r}_T^{(\tau_0)}(\tau),{\bf p}_T)
    \exp{\left(-\int\limits_{\tau_0}^\tau \tilde R_i(s,\theta^{(s)}(\tau),{\bf r}_T^{(s)}(\tau),{\bf p}_T)ds\right)}+ \nonumber\\
    &\int\limits_{\tau_0}^\tau d\lambda\left[
    f_i^{l.eq.}(\lambda,\theta^{(\lambda)}(\tau),{\bf r}_T^{(\lambda)}(\tau),{\bf p}_T)
    \tilde R_i(\lambda,\theta^{(\lambda)}(\tau),{\bf r}_T^{(\lambda)}(\tau),{\bf p}_T)+\tilde G_i^{decay}(\lambda,\theta^{(\lambda)}(\tau),{\bf r}_T^{(\lambda)}(\tau),{\bf p}_T) \right] \\
    &\exp\left(-\int\limits_{\lambda}^\tau \tilde R_i(s,\theta^{(s)}(\tau),{\bf r}_T^{(s)}(\tau),{\bf p}_T)ds\right) \nonumber
\end{align}
here $\tilde R_i(\lambda,\theta,{\bf r}_T,{\bf p}_T)=\frac{\cosh
y}{\cosh\theta} R_i(\lambda,\theta,{\bf r}_T,{\bf p}_T)$,
 $\tilde G_i^{decay}(\lambda,\theta,{\bf r}_T,{\bf p}_T)=\frac{\cosh y}{\cosh\theta} G_i^{decay}(\lambda,\theta,{\bf r}_T,{\bf p}_T)$.

Here we use the notation
\begin{equation}
\left\{\begin{array}{l}
\sinh\theta^{(\tau_0)}(\tau)=\frac{\tau}{\tau_0}\sinh\theta \\
{\bf r}_T^{(\tau_0)}(\tau)={\bf r}_T-\frac{{\bf p}_T}{m_T}(\tau\cosh\theta-\sqrt{\tau_0^2+\tau^2\sinh^2\theta})
\end{array}\right.
\end{equation}

The invariant value is $p^0 R_i(x,p)=p^{*0}R^*_i(x,p)$, where the
asterisk $^*$ denotes a value in the local rest frame of the fluid
element in point $x$, so
\begin{equation}
\tilde R_i(x,p)=\frac{\cosh y}{\cosh\theta} R_i(x,p)=\frac{\cosh
y}{\cosh\theta}\frac{p^\mu u_\mu}{p^0}R^*_i(p,T)= \frac{p^\mu
u_\mu}{m_T \cosh\theta}R^*_i(p,T)
\end{equation}

To connect the formal solution (\ref{f}) with observables, e.g.
particle spectrum, we use the equality
\begin{equation}
    p^0\frac{d^3n}{d^3p}=\frac{d^2n}{2\pi p_Tdp_Tdy}=\int\limits_{\sigma_{out}} d\sigma_\mu p^\mu f(x,p)
\end{equation}
where $\sigma_{out}$ is a "distant"\ hypersurface of large
$\tau=const$, where all the interactions among hadrons are ceased.

In what follows we use the variable substitution in the first term
of (\ref{f}) describing the "initial emission" :
\begin{equation}
    \left\{\begin{array}{l}
    \sinh\theta=\frac{\tau_0}{\tau}\sinh\theta' \\
    {\bf r}_T={\bf r}_T'+\frac{{\bf p}_T}{m_T}(\tau\cosh\theta-\sqrt{\tau_0^2+\tau^2\sinh^2\theta})
    \end{array}\right.
\end{equation}
and the substitution :
\begin{equation}
    \left\{\begin{array}{l}
    \sinh\theta=\frac{\lambda}{\tau}\sinh\theta' \\
    {\bf r}_T={\bf r}_T'+\frac{{\bf p}_T}{m_T}(\tau\cosh\theta-\sqrt{\lambda^2+\tau^2\sinh^2\theta})
    \end{array}\right.
\end{equation}
in the second term of (\ref{f}) related to the "4-volume emission".
After transformation to new variables $\{\tau,\theta',\vec r'\}$ we
arrive with the result:
\begin{align}\label{spectra}
    &\int\limits_{\sigma_{out}} d\sigma_\mu p^\mu f(x,p)=\int\limits_{\sigma_0} d\sigma_0^\mu p_\mu f_i^{l.eq.}(\tau_0,\theta',{\bf r}_T',p)
    \exp{\left(-\int\limits_{\tau_0}^\infty \tilde R_i(s,\theta^{(s)}(\tau_0),{\bf r}_T^{(s)}(\tau_0),{\bf p}_T)ds\right)}+ \nonumber\\
    &\int\limits_{\tau_0}^\tau d\lambda \int\limits_{\sigma(\lambda)}d\sigma_\mu(\lambda) p^\mu \left[f_i^{l.eq.}(\lambda,\theta',{\bf r}_T',{\bf p}_T)
    \tilde R_i(\lambda,\theta',{\bf r}_T',{\bf p}_T)+ \tilde G_i^{decay}(\lambda,\theta',{\bf r}_T',{\bf
p}_T)\right]\cdot \\
    &\cdot\exp\left(-\int\limits_{\lambda}^\infty \tilde
    R_i(s,\theta^{(s)}(\lambda),{\bf r}_T^{(s)}(\lambda),{\bf p}_T)ds\right)=p^0\frac{d^3N}{d^3p}\nonumber
\end{align}
where $\sigma(\lambda)$ is $\tau=\lambda=const$ hypersurface, so
$d\sigma_\mu(\lambda) p^\mu=\lambda m_T \cosh\theta' d\theta' d^2
\vec r_T'$. The exponential values in these expressions are the
escape probabilities
\begin{equation}
{\cal P}(\tau, {\bf r}_T, \theta, {\bf
p}_T)=\exp\left(-\int\limits_{\tau}^\infty \tilde
R_i(s,\theta^{(s)}(\tau),r_T^{(s)}(\tau),{\bf p}_T)ds\right)
\label{P}
\end{equation}
for particles with momentum $p$ at space-time point $(\tau, {\bf
r}_T, \eta=\theta+y)$ (in hyperbolic coordinates) to become free
without any collision \cite{PRL, PRC}.

In the expression above we can separate the 4-volume emission
function
\begin{equation}
    S_i(\lambda,\theta,{\bf r}_T,{\bf p}_T)=\left[f_i^{l.eq.}(\lambda,\theta,{\bf r}_T,p)
    \tilde R_i(\lambda,\theta,r_T,p)+ \tilde G_i^{decay}(\lambda,\theta,{\bf r}_T,{\bf p}_T)\right]
    {\cal P}(\lambda, {\bf r}_T,\theta,{\bf p}_T)
\end{equation}
and the initial emission :
\begin{equation}
    S_{i,0}(\theta,{\bf r}_T,{\bf p}_T)=f_i^{l.eq.}(\tau_0,\theta,{\bf r}_T,{\bf p}_T)
    {\cal P}(\tau_0, {\bf r}_T, \theta,{\bf p}_T)
\end{equation}

These expression demonstrate obviously that the particle emission is
formed by the particles which undergo their {\it last} interaction
or are already free initially. These expressions for the hadron
emission function are the basic functions for calculations of the
single- and multi- particle spectra \cite{PRL}. To evaluate these
quantities for observed (quasi) stable particles one needs to find
the term gain $G_i^{decay}$ for resonance decays and the collision
rates $R_i$.

\subsection{Resonance decays in multi-component gas} We suppose that
in the first (hydrodynamic) approximation the products of resonance
decays which interact with medium are thermalized and they become
free later, after the last collision with one of other particles.
However, at the late stages of matter evolution the system becomes
fairly dilute, so that some of these produced particles get a
possibility to escape without any collisions: ${\cal P}>0$. To
describe this we use the following form for $L_i^{decay}$ and
$G_i^{decay}$ terms (for 2-particle resonance decay) \cite{mrow}:
\begin{eqnarray}
p^0_iL_i^{decay}(x,p_i)=\sum_k\sum_l\int\frac{d^3p_k}{p^0_k}\int\frac{d^3p_l}{p^0_l}\Gamma_{i\rightarrow
kl}f_i(x,p_i)\frac{m_i}{F_{i\rightarrow kl}}
\delta^{(4)}(p_i-p_k-p_l)=m_i\Gamma_if_i(x,p_i) \label{loss}
\end{eqnarray}
where resonance $i$ decays into particles or resonances $k$ and $l$.
\begin{equation}\label{gdecay}
    p_i^0G_i^{decay}(x,p_i)=\sum_j\sum_k\int\frac{d^3p_j}{p^0_j}\int\frac{d^3p_k}{p^0_k}\Gamma_{j\rightarrow
ik}f_j(x,p_j)\frac{m_j}{F_{j\rightarrow ik}}
\delta^{(4)}(p_j-p_k-p_i)
\end{equation}
where the resonance $j$ decays into particles $i$ and $k$ with
partial width $\Gamma_{j\rightarrow ik}$ for this decay channel, and
\begin{equation}\label{j}
    F_{j\rightarrow ik}=
    \int\frac{d^3p_k}{p^0_k}\int\frac{d^3p_i}{p^0_i}\delta^{(4)}(p_j-p_k-p_i)=\frac{2\pi}{m_j^2}((m_j^2-m_k^2-m_i^2)^2-4m_i^2m_k^2)^{1/2}
\end{equation}

To escape the complicated problem with satisfaction of thermodynamic
identities in hadron resonance gas we utilize in what follows the
mass shell approximation for resonances, supposing that $m_i=\langle
m_i\rangle$. Also, as it was already discussed, we take into account
that the resonance mass in hadron resonance gas is much larger than
the temperature, $m_i >> T_c$. Then the most probable velocity of
resonance in the rest system of a fluid element is small,
$\overline{v}_i\approx \sqrt{\frac{2T}{m_i}}$, and one can use the
approximation
\begin{equation}\label{m}
p_i^{\mu}\approx m_i u^{\mu}.
\end{equation}
So the resonance distribution function takes the form
\begin{equation}
f_j(x,p_i)\approx \frac{p^0_j}{m_j}n_j(x)\delta^3({\bf p}_j-m_j{\bf
u}(x)),
\end{equation}
It allows us to perform integrations in (\ref{gdecay}) over $p_j$,
$p_k$ analytically and get :
\begin{equation}
G_i^{decay}(x,p_i)=\sum_j\sum_k \Gamma_{j\rightarrow
ik}\frac{n_j(x)}{p_i^0 p^0_k F_{j\rightarrow ik}}
\delta(m_ju^0(x)-p_k^0-p_i^0)
\end{equation}
where $p_k^0=\sqrt{m_k^2+(m_j{\bf u}(x)-{\bf p}_i)^2}$.

Just this form of gain term is used when spectra are evaluated
according to Eq. (\ref{spectra}). Note that in practical
calculations we substitute $\delta$-function by its Gaussian
representation:
$$\delta(x)=\frac{1}{R\sqrt\pi}e^{-x^2/R^2}$$ and take a finite parameter value $R=50 MeV$.

\subsection{Collision rates} The collision rate $R(x,p)=\frac{1}{\tau_{\text{rel}}(x,p)}$ is one of the basic value
for calculation of  the intensity of the interactions in the
expanding system and its decoupling. The latter is described through
the escape probability ${\cal P}(x,p)$ (\ref{P}) - the integral
value of $R$ along the possible trajectory of a particle with
momentum $p$ running freely through the whole expending system.  The
rate of collisions in the rest frame of some fluid element that
accounts for scatters of given particle with any other $i$th
hadronic species in the thermal Boltzmann system depends only on
particle energy $E^*_p=p^{\mu}u_{\mu}$ and the thermodynamic
parameters of this fluid element \cite{rate}:
\begin{eqnarray}
R^*(E^*_p, T,\{\mu_i\}) = \sum\limits_i  \int d^3k_i
\frac{g_{i}}{(2\pi)^3} \exp\left(-\frac{E_{k,i} -
\mu_{i}(x)}{T(x)}\right ) \sigma_{i}(s_i)
\frac{\sqrt{(s_i-(m-m_{i})^2)(s_i-(m+m_{i})^2)}}{2 E^*_{p}E_{k,i}}
\label{num-en5}
\end{eqnarray}
Here $g_{i}=(2j_{i}+1)$, $E_{p}=\sqrt{\mathbf{p}^{*2}+m^2}$,
$E_{k,i}=\sqrt{\mathbf{k}_i^2+m_{i}^2}$,  $s_i=(p^*+k_i)^{2}$ is the
squared c.m. energy of the pair, and $\sigma_{i}(s)$ is the total
cross section of selected particle with particle $i$ in the
corresponding binary collision. One can change the integration
variable to squared center of mass energy $s$, energy of scattering
partner $E_k$ and momentum angle $\phi$, and perform $E_k$ and
$\phi$-integration analytically, which gives the expression for
remaining integral:
\begin{align}
R^*(E^*_p, T,\{\mu_i\}) = & \sum_i\frac{g_iTe^{\mu_i/T}}{8\pi^2
p^*E^*_p}
\int\limits_{(m+m_i)^2}^{\infty}ds\sigma_i(s)\sqrt{(s-m^2-m^2_i)^2-4m_i^2 m^2}\times \nonumber\\
& \times\sinh\left(\frac{p*}{2Tm^2}\sqrt{(s-m^2-m^2_i)^2-4m_i^2
m^2}\right) \exp\left(-\frac{(s-m^2-m^2_i)E^*_p}{2Tm^2}\right)
\end{align}

We calculate $\sigma_{i}(s)$ in a way similar to UrQMD code
\cite{Bass}:
\begin{itemize}
 \item Breit-Wigner formula is applied for meson-meson and meson-baryon scattering:
\begin{eqnarray*}
\label{mbbreitwig} \sigma^{MB}_{total}(\sqrt{s}) &=&
\sum\limits_{R=\Delta,N^*}
       \langle j_B, m_B, j_{M}, m_{M} \| J_R, M_R \rangle \,
        \frac{2 S_R +1}{(2 S_B +1) (2 S_{M} +1 )}  \nonumber \\
&&\times        \frac{\pi}{p^2_{cm}}\,
        \frac{\Gamma_{R \rightarrow MB} \Gamma_{total}}
             {(M_R - \sqrt{s})^2 + {\Gamma_{tot}^2}/{4}}\quad,
\end{eqnarray*}
where $\Gamma_{total}=\sum\limits_{\text{\tiny (channels)}}\Gamma_{R
\rightarrow MB}$, with $\sqrt{s}$-dependent parametrization of
partial decay widths:
$$\Gamma_{R \rightarrow MB}(M)=\Gamma_R\frac{M_R}{M}\left(\frac{p_{CMS}(M)}{p_{CMS}(M_R)}\right)^{2l+1}
\frac{1.2}{1+0.2\left(\frac{p_{CMS}(M)}{p_{CMS}(M_R)}\right)^{2l}}$$
chosen to depend on absolute value  of particle momentum in
two-particle rest frame:
$$p_{CMS}(\sqrt{s})=\frac{1}{2\sqrt{s}}\sqrt{(s-m_1^2-m_2^2)^2-4m_1^2 m_2^2}$$
In the case of meson-meson scattering a constant elastic cross section of 5 mb is added in order to fully reproduce the measured cross section.

 \item PDG table data for $p-p$, $p-n$, $p-\bar p$, etc. scattering
 \item other baryon-baryon scattering: additive quark model:
$$
\sigma_{total} = 40\left(\frac{2}{3}\right)^{m_1+m_2}
\left(1-0.4\frac{s_1}{3-m_1}\right) \left(1-0.4\frac{s_2}{3-m_2}
\right) \, [{\rm mb}]\quad,
$$
$m_i =1 (0)$ corresponds to meson(baryon), $s_i$ - number of strange
quarks in hadron $i$.
\end{itemize}
Note that all relevant resonance states (see above), $359$ different species - are taken into account
for the calculation of $\sigma_{i}(s)$.

\section{Hydrodynamics}

We describe the system evolution in the equilibrium zone at $T >
T_{ch}$  by the perfect hydrodynamics. The small shear viscosity
effects, which lead to an increase of the transverse flows
\cite{Teaney} we account phenomenologically including this effect in
the parameter $\alpha$ of initial velocity as described in Section
IIA. The matter evolution in this zone is described by the
relativistic hydrodynamical equations related to the conservation of
energy-momentum:
\begin{equation}
 \partial_\nu T^{\mu\nu}=0
 \label{hydro_eq}
\end{equation}
and equations associated with the net baryon number, strangeness and
isospin conservations
\begin{equation}
\partial_\nu(q_i u^\nu)=0
\end{equation}
Here $q_i$ is the density of conserved quantum number.

At $T < T_{ch}$ the equations for the system evolution in the first
approximation: $f_i=f_i^{l.eq}$ can be derived from the basic
equation (\ref{rel}). Namely, integrating the left and right hand
sides of Eq.(\ref{rel}) over $d^3p$ one arrives to the equation
(\ref{decay}) for particle number flow in the non-equilibrium zone,
and also to hydrodynamic equation (\ref{hydro_eq}) by integrating
Eq. (\ref{rel}) over $p_i^{\nu}d^3p_i$ and summing over index $i$.

Note that in the first approximation the matter evolution is
described by the equations of ideal hydrodynamics while the
distribution function (\ref{f}) in decaying system is
non-equilibrium. The iteration procedure, including the next order
approximations, that, in fact, leads to viscous hydrodynamic
evolution, is described in Ref. \cite{PRC}. In this article we limit
ourself by the first approximation. Then the energy-momentum tensor
$T^{\mu\nu}$ has a simple structure which is employed in this model
:
\begin{equation}
T^{\mu\nu}=(\epsilon+p)u^\mu u^\nu - p\cdot g^{\mu\nu},
\end{equation}
where $\epsilon$ is energy density and $p$ is pressure  defined from
the EoS. In the chemically equilibrated zone the pressure is defined
from the lattice QCD calculations as discussed in Section IIIA. In
the non-equilibrium zone the EoS generally depends on all 360
variables, $p=p(\epsilon, \{n_i\})$, and it is evaluated altogether
with solution of the evolutionary  equations. The reason is that it
is impossible to store the EoS table, therefore we compute pressure
each time we need it (e.g. when restoring thermodynamic variables
from conservative variables or when computing fluxes through each
cell boundary) solving analytic equations (\ref{Tboltz}),
(\ref{pboltz}) numerically.

Let us  rewrite equations in hyperbolic coordinates. These
coordinates are suitable for dynamical description at RHIC energies,
since, for example, zero longitudinal flows correspond to
boost-invariant expansion (so nonzero longitudinal flow correspond
to deviation from boost invariance), and evolution parameter,
$\tau=\sqrt{t^2-x^2}$ is not affected by strong longitudinal flow,
which saves computational time. It is convenient to write the
equations in conservative form, then the conservative variables are
:
\begin{equation}
    \vec Q=
    \left(\begin{array}{c}
    Q_\tau \\
    Q_x \\
    Q_y \\
    Q_\eta \\
    \{ Q_{n_i} \}
    \end{array}\right)=
    \left(\begin{array}{c}
    \gamma^2(\epsilon+p)-p \\
    \gamma^2(\epsilon+p)v_x \\
    \gamma^2(\epsilon+p)v_y \\
    \gamma^2(\epsilon+p)v_\eta \\
    \{ \gamma n_i \}
    \end{array}\right).
\end{equation}
Here the expression in curly brackets denote $N$ variables
associated with the particle densities for each sort of hadrons. The
$Q_i$ are conservative variables in the sense that integral
(discrete sum over all cells) of $Q_i$ gives the total energy,
momentum and particle numbers, which are conserved up to the fluxes
on the grid boundaries. The velocities in this expression are
defined in LCMS (longitudinally co-moving system), and related to
velocities in the laboratory frame as :
\begin{align}
v_x&=v_x^{lab}\cdot\frac{\cosh y_f}{\cosh(y_f-\eta)} \nonumber\\
v_y&=v_y^{lab}\cdot\frac{\cosh y_f}{\cosh(y_f-\eta)} \nonumber\\
v_\eta&=\th(y_f-\eta),
\end{align}
where $y_f=\frac 1 2 \ln[(1+v^{lab}_z)/(1-v^{lab}_z)]$ is the
longitudinal rapidity of fluid element, $\eta=\frac 1 2
\ln[(t+z)/(t-z)]$ is the geometrical rapidity.

The full hydrodynamical equations are :
\begin{equation}
\partial_\tau\underbrace{
\left(\begin{array}{c}
    Q_\tau\\
    Q_x \\
    Q_y \\
    Q_\eta \\
    \{ Q_{n_i} \}
    \end{array}\right)}_{\text{quantities}}+
    \vec\nabla\cdot\underbrace{\left(\begin{array}{c}
    Q_\tau\\
    Q_x \\
    Q_y \\
    Q_\eta \\
    \{ Q_{n_i} \}
    \end{array}\right)\vec v +
    \left(\begin{array}{c}
    \vec\nabla(p\cdot \vec v)\\
    \partial_x p \\
    \partial_y p \\
    \frac 1 \tau \partial_\eta p \\
    0
    \end{array}\right)}_{\text{fluxes}}+
    \underbrace{
    \left(\begin{array}{c}
    (Q_\tau+p)(1+v_\eta^2)/\tau\\
    Q_x/\tau \\
    Q_y/\tau \\
    2Q_\eta/\tau \\
    \{ Q_{n_i}/\tau \}
    \end{array}\right)}_{\text{sources}}
    =0
\end{equation}
and $\vec\nabla=\left(\partial_x,\ \partial_y,\ \frac 1
\tau\partial_\eta\right)$.

For hydrodynamic calculations related to midrapidity region on
central A+A collisions we impose longitudinal symmetry and
cylindrical symmetry in transverse direction. This actually means
that tangential (in transverse direction) and longitudinal
velocities in LCMS vanish, so $Q_\phi=Q_\eta=0$, as well as the
fluxes in $\phi$ and $\eta$ directions. Then, one has to solve the
following set of equations :
\begin{equation}
\partial_\tau\underbrace{
\left(\begin{array}{c}
    Q_\tau\\
    Q_r \\
    \{ Q_{n_i} \}
    \end{array}\right)}_{\text{quantities}}+
    \partial_r\cdot\underbrace{\left(\begin{array}{c}
    (Q_\tau+p)v_r\\
    Q_r v_r+p \\
    \{ Q_{n_i}v_r \}
    \end{array}\right)}_{\text{fluxes}} +
    \underbrace{
    \left(\begin{array}{c}
    (Q_\tau+p)(1+v_\eta^2)/\tau-(Q_\tau+p)v_r/r\\
    Q_r/\tau-Q_r v_r/r \\
    \{ Q_{n_i}/\tau-Q_{n_i}v_r/r \}
    \end{array}\right)}_{\text{sources}}
    =0
\end{equation}
Practically $v_r/r$ is ambiguous at $r=0$, so we put $v_r/r=\alpha$
there and use $\alpha$ value interpolated from the neighboring
points.

We base our calculations on the finite-volume approach: we
discretize the system on a fixed grid in the calculational frame and
interpret $Q_i^n$ as average value over some space interval $i$,
which is called a cell ($i$ is a multi-index in multidimensional
case). We also split continuous time evolution into a sequence of
finite timesteps $n$.

The $Q_i^n$ are then updated after each timestep according to the
fluxes on the cell interface during the timestep $\Delta t_n$. In
3-dimensional case one has the following update formula :
\begin{equation}
Q_{ijk}^{n+1}=Q_{ijk}^n-\frac{\Delta t}{\Delta
x_1}(F_{i+1/2,jk}+F_{i-1/2,jk})-\frac{\Delta t}{\Delta
x_2}(F_{i,j+1/2,k}+F_{i,j-1/2,k})-\frac{\Delta t}{\Delta
x_3}(F_{ij,k+1/2}+F_{ij,k-1/2})
\end{equation}
Where $F$ is the average flux over the cell boundary, indexes $+1/2$
and $-1/2$ correspond to right and left cell boundary in each
direction.

This is a base of the Godunov method \cite{Holt} that implies that
the distributions of variables on the grid are piecewise, this forms
the Riemann problem at each cell interface. Then the flux through
each cell interface depends only on the solution of single Riemann
problem, supposing that the waves from the neighbouring
discontinuities do not intersect. The latter is satisfied with
Courant-Friedrichs-Lewy (CFL) condition \cite{CFL}.

To solve the Riemann problems at each cell interface we use
relativistic HLLE solver \cite{Schneider}, which approximates the
wave profile in the Riemann problem by single intermediate state
between two shock waves propagating away from the initial
discontinuity. Together with the shock wave velocity estimate, in
this approximation one can obtain analytical dependence of flux on
initial conditions for Riemann problem, which makes algorithm to be
explicit.

We proceed then to construct higher-order numerical scheme. To do
so,
\begin{itemize}
    \item in time : the \emph{predictor-corrector} scheme is used for the second order of accuracy in time, i.e. the numerical error is $O(dt^3)$, instead of $O(dt^2)$
    \item in space : in the same way, to achieve the second order scheme the \emph{linear distributions} of quantities (conservative variables) inside cells are used.
\end{itemize}
--- {\it Multi-dimension problem.} At each timestep, we compute and
sum the fluxes for each cell with all its neighbours and update the
value of conservative variables with the total flux. Thus, we do not
use operator splitting (dimensional splitting) and thus avoid the
numerical artifacts introduced by this method, e.g. artificial
spatial asymmetry.

--- {\it Grid boundaries.} To treat grid boundaries, we use the method
of \emph{ghost cells}. We include the two additional cells on either
end of grid in each direction, and set the quantities in these cells
at the beginning of each timestep. For simplicity, we set the
quantities in ghost cells to be equal to these in the nearest "real"
cell, thus implementing nonreflecting boundary conditions (outflow
boundary). This physically correspond to boundary which does not
reflect any wave, which is consistent with expansion into vacuum.

--- {\it Vacuum treatment.} In our simulations we deal with spatially
finite systems expanding into vacuum. Thus computational grid in the
Eulerian algorithm must initially contain both system and
surrounding vacuum. To account for a finite velocity of expansion
into vacuum, which equals $c$ for infinitesimal slice of matter on
the boundary, we introduce additional (floating-point) variables in
each cell which keep the extent of matter expansion within a cell,
having value 1 for the complete cell, 0 for a cell with vacuum only.
The matter is allowed to expand in the next vacuum cell only if the
current cell is filled with the matter.

\section{Results and discussion}
In the article we apply the hydrokinetic model for an analysis of
the space-time picture of \textrm{Au}+\textrm{Au} collisions at the
top RHIC energies. Such an analysis provided in the evolutionary
models of heavy ion collisions  have to be based on a detailed
description of the pion and kaon femtoscopic scales and must also
describe well the {\it absolute} values of the spectra (not only
spectra slopes) of the particles.  As it was noted in Ref.
\cite{PRC}, the following factors favor the simultaneous description
of the mentioned data: a relatively hard EoS (crossover transition
between hadronic and quark-gluon matters, not the first order phase
transition), the pre-thermal transverse flows developed to
thermalization time, an account for an "additional portion" of the
transverse flows due to the shear viscosity effect \cite{Teaney}, a
correct description of the gradual decay of the system at the late
stage of the expansion. All these factors are included in the
presented version of the HKM.

We use both the Glauber-like (Section IIB) and CGC-like (Section
IIC) initial conditions. In the former case the mean transverse
radii, defined by (\ref{yT}) is $R_T = 4.137$ fm for the top RHIC
energy. The best fit for the Glauber IC is reached at the following
values of the two fitting parameters related to the proper time
$\tau = 1$ fm/c: $\epsilon_0 = 16.5$ GeV/fm$^3$ ($\langle \epsilon
\rangle = 11.69$ GeV/fm$^3$) and parameter of the initial velocity
defined by (\ref{yT}), $\alpha$ = 0.248 ($\langle v_T \rangle =
0.224$). In the case of the CGC-like initial conditions $R_T = 3.88$
fm, the fitting parameters leading to the best data description are
$\epsilon_0 = 19.5$ GeV/fm$^3$ ($\langle \epsilon \rangle = 13.22$
GeV/fm$^3$) and $\alpha$ = 0.23 ($\langle v_T \rangle = 0.208$). The
parameters $\alpha$ for the initial transverse flows are somewhat
larger than they are for the free streaming approximation of the
pre-thermal stage \cite{Sin-2}. The reason is, as it is explained in
Section II, that the fitting parameter $\alpha$ is related to the
"unknown portions"  of flows, caused by the two factors: a
developing of the pre-thermal flows and the viscosity effects in the
quark-gluon plasma. In addition, an account of the event-by-event
fluctuations of the initial conditions also leads to an increase of
the "effective" transverse flows, obtained by averaging at the final
stage, as compare to the results based on the initial conditions
averaged over initial fluctuations \cite{Hama1}. Since we use the
later kind of IC, it should  lead also to an increase of the
effective parameter $\alpha$.

\begin{figure}
\hspace{0cm}
\includegraphics[scale=0.4]{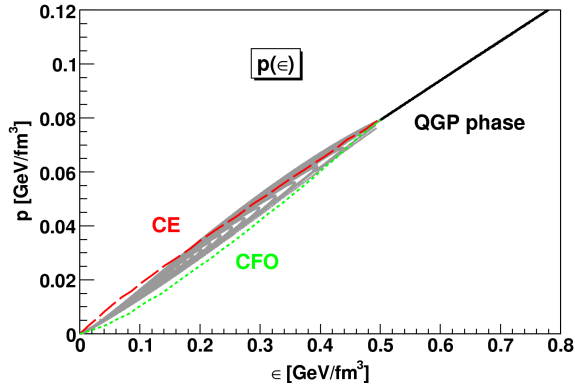}
\caption{(Color online) Equation of state $p(\epsilon)$ used in the HKM
calculations. The solid black line is related to the chemically
equilibrated phase, taken from lattice QCD results as described in
Sec. IIIA, while grey region consists of set of the points
corresponding to the different hadron gas compositions at each
$\epsilon$ occurring during the late non-equilibrium stage of the
evolution. The dashed line denotes EoS for the chemically
equilibrated hadron gas and dotted line for the chemically frozen
one, they are shown for a comparison.}
\end{figure}

As it was discussed in Sections III, the chemically non-equilibrated
evolution at the late stage, $T<T_{ch}=165$ MeV, is not
characterized by a simple EoS, like $p=p(\epsilon,\mu_B)$, in our
calculations the pressure in this domain depends on 360 variables:
energy density and particle concentrations. In Fig. 1 we demonstrate
the "effective" EoS at the temperatures around and below $T_{ch}$.
The points related to the later region characterize all diapason
which the pressure gains at each energy density when the system
evolves with the Glauber IC fixed above. We see that the pressure is
differ from the "limited" cases: the chemically equilibrated and
completely chemically frozen evolution (when the numbers of {\it
all} (quasi) stable particles and resonances are conserved). At
relatively large energy densities in a dominated space-time region
the non-equilibrium EoS is harder than even in the chemically
equilibrated case. This could reduce the out- to side- ratio for
transverse interferometry radii.

The results of the HKM for the pion and kaon spectra, interferometry
radii and $R_{out}/R_{side}$ ratio are presented in Fig. 2. Since
the temperature and baryonic chemical potential at chemical
freeze-out, which are taken from the analysis of the particle number
ratios \cite{Becattini}, are is more suitable for the STAR
experiment, the HKM results for kaon spectra are good for the STAR
data but not so much for the PHENIX ones. Note also that, in spite
of other studies (e.g., \cite{Broniowski}), we compare our results
for the interferometry radii within the whole measured interval of
$p_T$ covered at the top RHIC energy. Finally, one can conclude from
Fig. 2 that the description of pion and kaon spectra and space-time
scales is quite good for both IC, the Glauber and CGC. It is worth
noting, however, that the two fitting parameters $\alpha$ and
$\epsilon_0$ are various by 10-20$\%$ for different IC, as it is
described above.

The special attention acquires a good description of the pion and
kaon longitudinal radii altogether with $R_{out}/R_{side}$  ratio,
practically, within the experimental errors. Such an achievement
means that the HKM catches the main features of the matter evolution
in A+A collisions and correctly reproduces the homogeneity lengths
in the different parts of the system which are directly related to
the interferometry radii at the different momenta of the pairs
\cite{Sin}. In this connection it is valuable to show the structure
of the emission function for pions and kaons.

In Fig. 3 we demonstrate the space-time structure of the particle
emission at the Glauber IC for different transverse momenta of
particles, longitudinal momenta is close to zero. The space-time
picture of particle liberation is quite different for different
transverse momenta: for the soft particles the maximal emission
occurs close to the cental part and happens at relatively later
times, while the most of the hard particles are emitted from the
periphery of the system at early times. In fact (see also \cite{PRC,
sin-3}), the temperatures in the regions of the maximal emission are
quite different for different $p_T$, they are for pions: $T\approx
75-110$ MeV for $p_T=$ 0.2 GeV/c and $T\approx 130-135 $ MeV for
$p_T=$ 1.2 GeV/c. So, if one uses the generalized Cooper-Frye
prescription \cite{PRC, sin-3} applied to the {\it hypersurfaces of
the  maximal emission}, these hypersurfaces will be different for
the different particle momenta and does not correspond to common
isotherm \cite{PRC, sin-3}.

One can see in Fig. 3, the top plots, that at {\it equal} transverse
momentum $p_T$ the maximal emission of kaons happens earlier than
pions as one can expect since the kaons interact weaker. At the same
time the kaon interferometry radii in Fig. 2 follow approximately to
the pion radii, demonstrating the approximate $m_T$-scaling
\cite{Averch} with deviations to the slightly bigger values than
pion radii have. The explanations can be gain from the middle row in
Fig. 3 where the comparison is done for the same transverse mass of
pions and kaons. Then the maxima of pion and kaon emissions become
closer and the majority of kaons leave system even somewhat later
than pions at the same $m_T$, opposite to the comparison at the same
$p_T$. Since in simplest situations the homogeneity lengths for
bosons depend on $m_T$ \cite{Averch}, one could say that the
approximate $m_T$-scaling could indicate the similarity of the
freeze-out picture for kaons and pions.  However, probably, such a
conclusion is very approximate since the real structure of the
emission processes in A+A collisions is quite complicated as one can
see from the details in Fig.3.

\begin{figure*}
\hspace{0cm}
\includegraphics[scale=0.8]{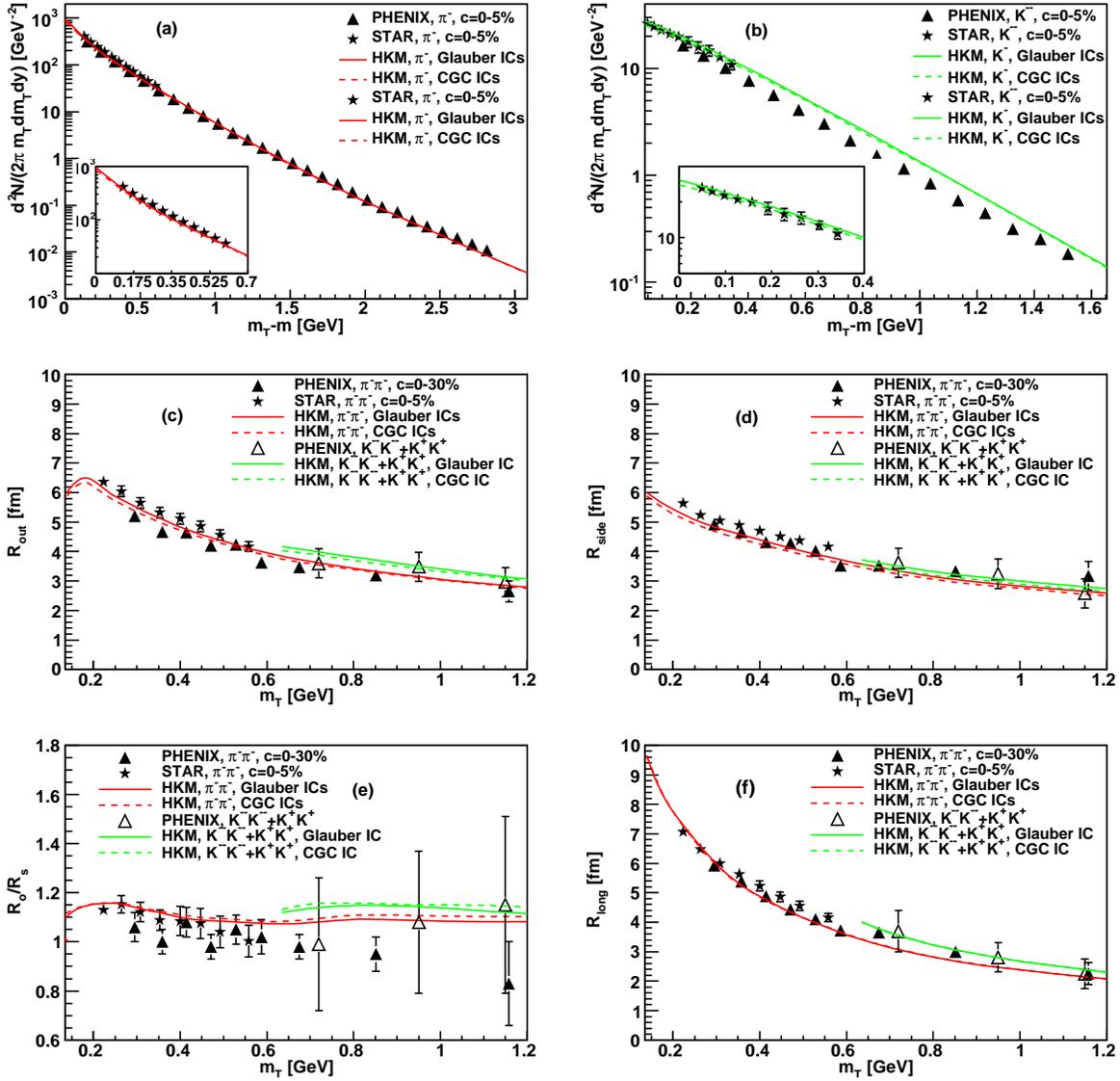}
\caption{(Color online) The transverse momentum spectra of (a) negative pions and
(b) negative kaons, all calculated in the HKM model. The comparison only
with the STAR data are presented in the separate small plots.
The interferometry radii: (c) $R_{out}$, (d) $R_{side}$, (f) $R_{long}$ and (e)~$R_{out}/R_{side}$ ratio for
$\pi^{-}\pi^{-}$ pairs and mixture of $K^- K^-$ and $K^+ K^+$ pairs.
The experimental data are taken from the STAR
\cite{star-spectra, star-hbt} and PHENIX \cite{phenix-spectra,
phenix-hbt, phenix-hbt-kaon} Collaborations.}
\end{figure*}

\begin{figure*}
\hspace{0cm}
\includegraphics[scale=0.55]{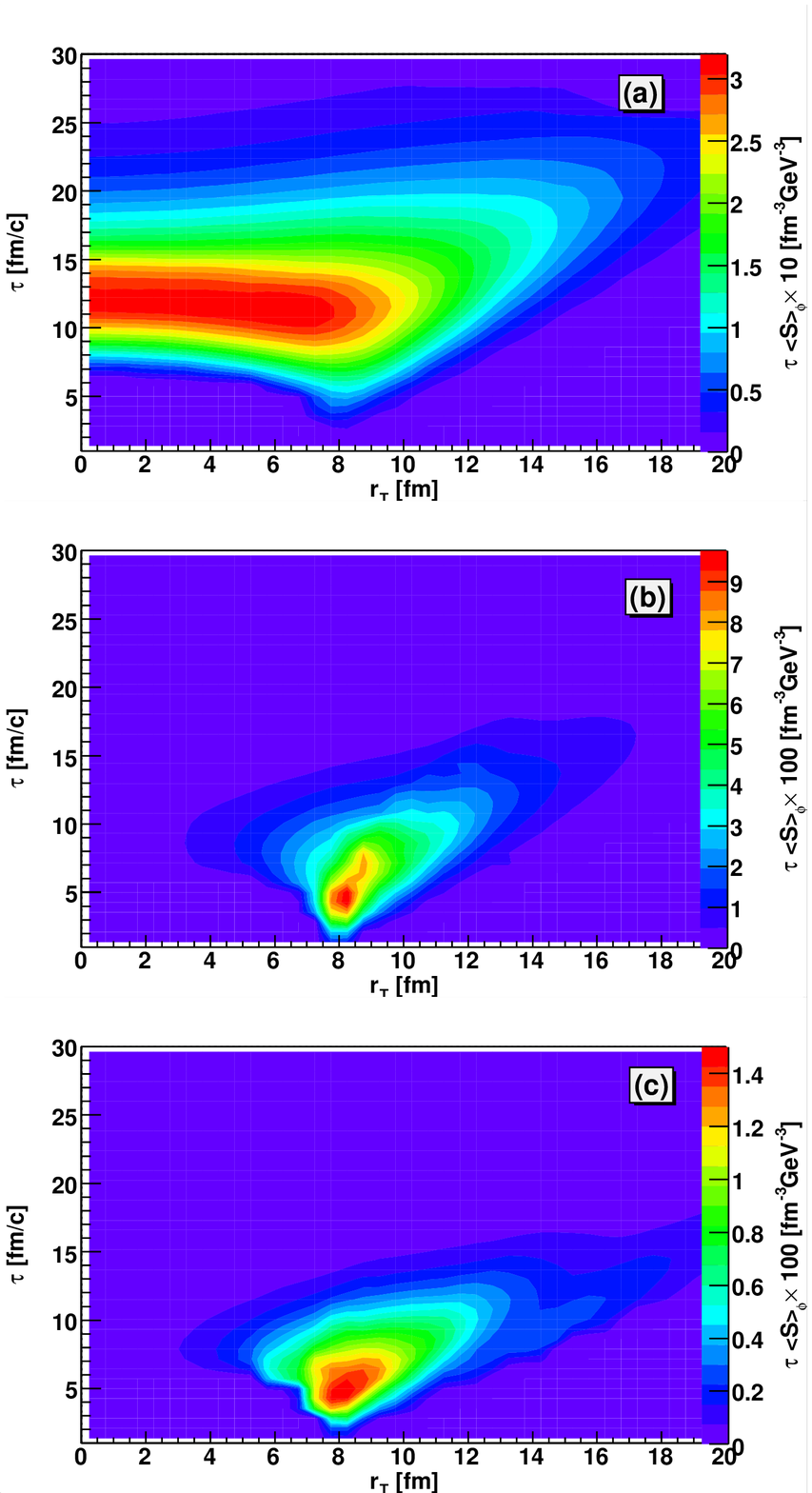}
\includegraphics[scale=0.55]{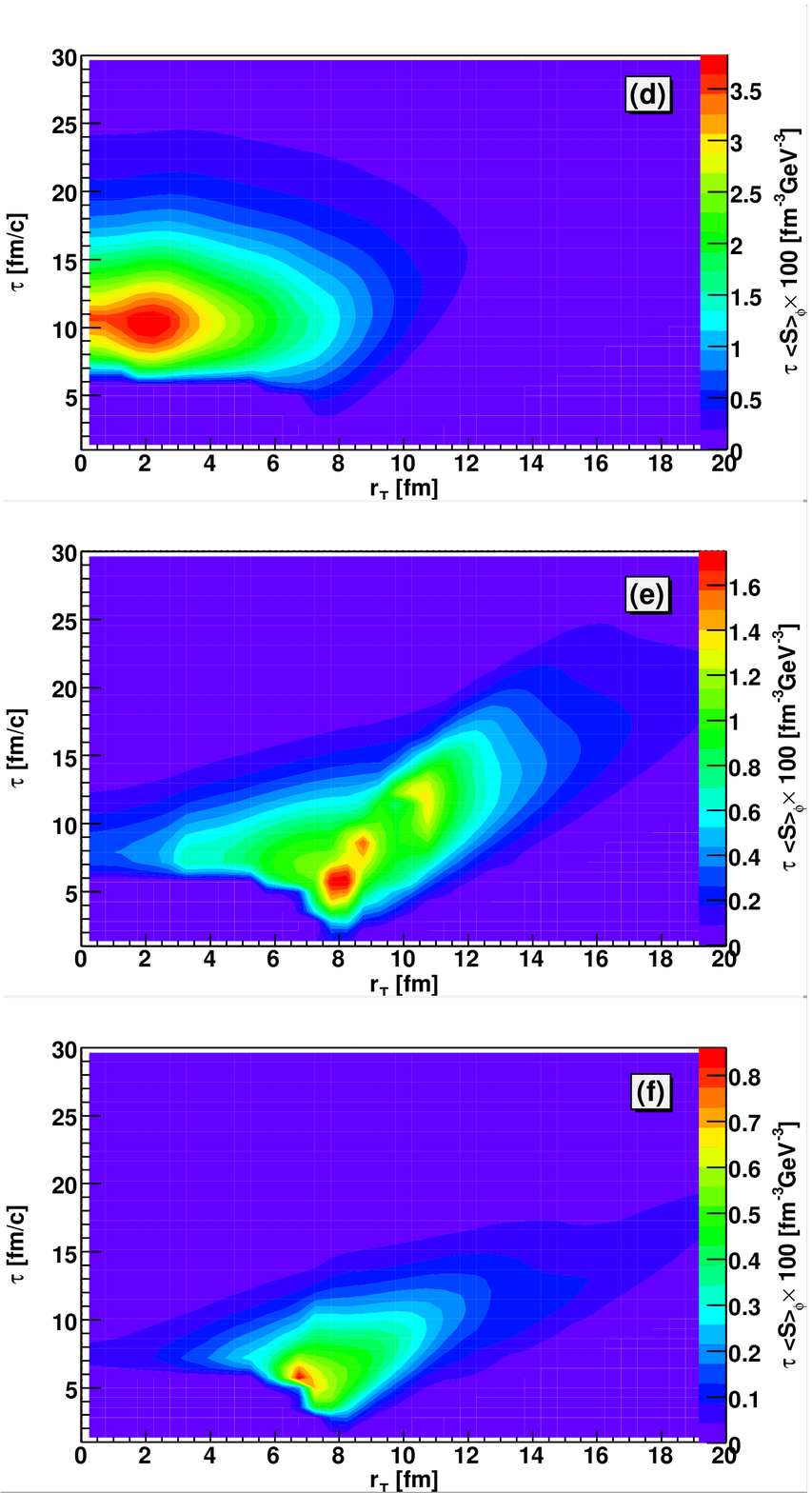}
\caption{(Color online) The $\phi_p$-integrated emission functions of (a,b,c) negative
pions and (d,e,f) negative kaons with different momenta: (a,d) $p_T=0.2$ GeV, (b) $p_T=0.85 GeV$, (e) $p_T=0.7 GeV$, (c,f) $p_T=1.2 GeV$ at the Glauber IC.
The values of $p_T$ in the middle row (b,e) correspond to the same
transverse mass for pions and kaons $m_T=0.86$ GeV.}
\end{figure*}

\section{Conclusions}

The hydro-kinetic model \cite{PRL, PRC} is developed  for a detailed
study of the matter evolution and space-time picture of hadronic
emission from rapidly expanding fireballs in \textit{A}+\textit{A}
collisions. The model allows one to describe the evolution of the
QGP as well as the gradually decoupling hadronic fluid - a
chemically non-equilibrium matter, where the equation of state is
defined  at each space-time point and accounts for decays of
resonances into the non-equilibrated medium.

The HKM is applied to restore the initial conditions and space-time
picture of the matter evolution in central Au+Au collisions at the
top RHIC energy. The analysis, which is based on a detailed
reproduction of the pion and kaon momentum spectra and measured
femtoscopic scales, demonstrates that basically the pictures of the
matter evolution and particle emission are similar at both Glauber
and CGC initial conditions (IC) with, however, the different initial
maximal energy densities: it is about 20\% more for the CGC initial
conditions. The initial pre-thermal flow is slightly less for the
CGC IC. The main factors, which allows one to describe well
simultaneously the spectra and femtoscopic scales are: a relatively
hard EoS (crossover transition and chemically non-equilibrium
composition of hadronic matter), pre-thermal transverse flows
developed to thermalization time, an account for an "additional
portion" of the transverse flows due to the shear viscosity effect
and fluctuation of initial conditions, a correct description of a
gradual decay of the non-equilibrium fluid at the late stage of
expansion. Then one does not require the too early thermalization
time, $\tau_i<1$ fm/c, to describe the data well. All these factors
are included in the presented version of the HKM and it allows one
to describe observables with only the two parameters.

An analysis of the emission function at the top RHIC energies
demonstrates that the process of decoupling of the fireballs created
in Au+Au collision lasts from about 8 to 20 fm/c, more than the half
of fireball's total lifetime. The temperatures in the regions of the
maximal emission are different at the different transverse momenta
of emitting pions: $T\approx 75-110$ MeV for $p_T=$ 0.2 GeV/c and
$T\approx 130-135 $ MeV for $p_T=$ 1.2 GeV/c. A comparison of the
pion and kaon emissions at the same transverse mass demonstrates the
similarity of the positions of emission maxima, that could point out
to the reason for an approximate $m_T$ scaling.

Summary: the advanced HKM tool allows one to describe the process of
the fireball evolution and gradual particle liberation in agreement
with underlying kinetic equations.  Further developments of the
hydrokinetic approach and an analysis of the data in non-central A+A
collisions will be the subject of a follow-up work.

\section*{Acknowledgments}

The authors thank S.V. Akkelin for fruitful discussions. The
research was carried out within the scope of the EUREA: European
Ultra Relativistic Energies Agreement (European Research Group:
Heavy ions at ultrarelativistic energies) and is supported by the
Fundamental Researches State Fund of Ukraine, Agreement No
F33/461-2009 with Ministry for Education and Science of Ukraine. In
part it was supported also within the Ukrainian-Russian grant,
Agreement No $\Phi$28/335-2009.

\end{document}